\documentclass[apj]{emulateapj}
\usepackage{graphicx}
\usepackage{amsmath,amssymb}
\usepackage{amsmath}

\usepackage[normalem]{ulem}
\usepackage[usenames]{color}

\def\mnras{MNRAS}
\def\apj{ApJ}

\let\oldAA\AA
\renewcommand{\AA}{\text{\normalfont\oldAA}}

\begin{document}

\title{Antlia2's role in driving the ripples in the outer gas disk of the Galaxy}

\author {Sukanya Chakrabarti  \altaffilmark{1}, 
Philip Chang  \altaffilmark{2,3},
Adrian M. Price-Whelan \altaffilmark{4}, 
Justin Read \altaffilmark{5}, 
Leo Blitz  \altaffilmark{6}, 
Lars Hernquist \altaffilmark{7}}

\altaffiltext{1}
{School of Physics and Astronomy, Rochester Institute of Technology, 84 Lomb Memorial Drive, Rochester, NY 14623; chakrabarti@astro.rit.edu}
\altaffiltext{2}
{Department of Physics, University of Wisconsin-Milwaukee, 3135 North Maryland Avenue, Milwaukee, WI 53211}
\altaffiltext{3}
{Center for Computational Astrophysics, Flatiron Institute, 162 Fifth Avenue, New York, NY, 10010}
\altaffiltext{4}
{Princeton University}
\altaffiltext{5}{University of Surrey}
\altaffiltext{6}{UC Berkeley}
\altaffiltext{7}{Harvard University}

\begin{abstract}
We employ the earlier published proper motions of the newly discovered Antlia 2 dwarf galaxy derived from Gaia data to calculate its orbital distribution in the cosmologically recent past.  Using these observationally motivated orbits, we calculate the effect of the Antlia 2 dwarf galaxy on the outer HI disk of the Milky Way,  using both test particle and Smoothed Particle Hydrodynamics simulations.  We find that orbits with low pericenters, $\sim$ 10 kpc, produce disturbances that match the observed outer HI disk perturbations.  We have independently recalculated the proper motion of the Antlia 2 dwarf from Gaia data and found a proper motion of $(\mu_{\alpha}cos\delta, \mu_{\delta}) = (-0.068,0.032) \pm (0.023,-0.031)~\rm mas/yr$, which agrees with results from Torrealba et al. (2019) within the errors, but gives lower mean pericenters, e.g., $\sim$ 15 kpc for our fiducial model of the Milky Way.  We also show that the Sagittarius dwarf galaxy interaction does not match the observed perturbations in the outer gas disk.  Thus, Antlia 2 may be the driver of the observed large perturbations in the outer gas disk of the Galaxy.  The current location of the Antlia 2 dwarf galaxy closely matches that predicted by an earlier dynamical analysis (Chakrabarti \& Blitz 2009) of the dwarf that drove ripples in the outer Galaxy, and, in particular, its orbit is nearly coplanar to the Galactic disk.  If the Antlia 2 dwarf galaxy is responsible for the perturbations in the outer Galactic disk, it would have a specific range of proper motions that we predict here; this can be tested soon with Gaia DR-3 and Gaia DR-4 data.  
\end{abstract}

\section{Introduction}
The recently discovered Antlia 2 dwarf galaxy (Torrealba et al. 2019) is unique in several ways.  At a distance of $\sim$ 130 kpc, and a half-light radius of 2.9 kpc (similar in extent to the Large Magellanic Cloud, but two magnitudes fainter), it is the lowest surface brightness system known.  Fritz et al. (2018) have noted that the fact that there are fewer dwarf galaxies observed near apocenters vs near pericenters suggests that there are more dwarf galaxies to be discovered.  Possibly, Antlia 2 falls into this group of dwarfs close to apocenters that are just being discovered.  Yet another intriguing aspect of Antlia 2 is that with a mean [Fe/H] metallicity of -1.4, its inferred stellar mass from the mass-metallicity relation (Kirby et al. 2013) would be $\sim 10^{7} M_{\odot}$.  However, substantially lower values result from its current luminosity assuming standard mass-to-light ratios (Torrealba et al. 2019), which would suggest that it has undergone significant tidal disruption. 

The planar disturbances manifest in the outer HI disk of the Milky Way (Levine, Blitz \& Heiles 2006; henceforth LBH06) have been a long standing puzzle.  Chakrabarti \& Blitz (2009; henceforth CB09) analyzed the perturbations observed in the outer HI disk of the Milky Way (LBH06).  They argued that a new dwarf galaxy was needed to explain the observed disturbances and predicted its orbital parameters.  Namely, they found that the observed outer disk planar disturbances could be explained by a $\sim$ 1:100 mass ratio perturber on a near co-planar orbit with a close pericenter approach ($R_{\rm peri} \sim 5~h^{-1}$ kpc) that is currently at a distance of $\sim$ 90~$h^{-1}$ kpc, where the small pericenter and co-planar orbit is constrained by the strength of the observed disturbances and the current distance by the timescale for the initial orbital perturbations to manifest itself as surface density perturbations. 

Here, we use the observed Gaia proper motions of the Antlia 2 dwarf galaxy to investigate if the Antlia 2 dwarf galaxy can produce the observed disturbances in the outer HI disk of the Milky Way.  We use test particle calculations (Chang \& Chakrabarti 2011; henceforth CC11) and fitting relations (Lipnicky, Chakrabarti \& Chang 2018; henceforth LCC18) to survey the parameter space, to determine the approximate response.  We then carry out a smaller set of targeted SPH calculations with GADGET-2 (Springel 2005).  We find that the low pericenters $R_{\rm peri} \sim 10$ kpc of the orbital distribution can explain the observed disturbances in the outer HI disk. The tidal debris of the Sgr dwarf suggests that it has approached relatively close to the Galactic disk (Newberg et al. 2003), and models (Purcell et al. 2011; Laporte et al. 2017; D'Onghia et al. 2016; Haines et al. 2019) have suggested that it has excited various features in the Galactic disk.   We show here however that it is not responsible for the large planar disturbances in the outer HI disk.

Analysis of recent observations (Fritz et al. 2018), and of recent cosmological simulations (Garrison-Kimmel et al. 2018; Samuel et al. 2019) suggests that there may be one (or more) dwarf galaxies now at apocenter, that suffered close approaches to the Galaxy.  If correct, the perturbation that such a dwarf galaxy would exert on the Galactic disk ought to be explored.  

This paper is organized as follows.  In \S 2, we review our methodology for test particle and SPH simulations.  In \S 3, we first give our results from orbit integrations for the pericenter distributions where we employ the proper motions reported in Torrealba et (2019).   We also discuss the results from test particle calculations and the SPH simulations for the Fourier amplitudes and the phase.  In \S 3.1, we employ all the Antlia 2 stars and recalculate the proper motions and the corresponding pericenter distributions, which results in the lower pericenters becoming more probable.  In \S 4, we discuss future work and conclude. Specifically, we include here a prediction of the proper motions that Antlia2 should have if it is in fact the dwarf galaxy that perturbed the Milky Way, which can be tested in the near future by Gaia DR-3 and DR-4 measurements which will have significantly lower errors in the proper motion measurements.

\section{Methodology}

\begin{table*}
\centering
        \caption{\textbf{Simulation Parameters}}

          \begin{tabular}{@{}lccc@{}}
          \hline

Simulation Name	 &  $V_{200}$ [km/s] &  $M_{200}$ [$M_{\odot}$] & $R_{\rm peri}$   \\	
\hline

FRp8 &  200 &  $1.8\times 10^{12} $ &  8                  \\
FRp16 &  200 &  $1.8\times 10^{12} $ &  16                \\
FRp32 &  200 &  $1.8\times 10^{12} $ &  32                \\
Sgr    & 200   & $1.8 \times 10^{12}$ & 13         \\
GRp8 &  180 &  $1.3\times 10^{12} $ &  10               \\
GRp16 & 180 &  $1.3\times 10^{12} $ &  16                \\
GRp32 & 180 &  $1.3\times 10^{12} $ &  32             \\ 
\hline

\end{tabular} 
\end{table*} 

We first integrate the orbits and sample the errors in the observed Gaia proper-motions as reported in Torrealba et al. (2019), to determine an orbital distribution for the Antlia 2 dwarf galaxy.  In \S 3.1, we employ all the Antlia 2 stars and re-calculate the proper motions and associated pericenter distributions.  The orbits of Antlia 2 are integrated backwards in time in a Hernquist (1990) potential that is matched to the Navarro, Frenk \& White (1996) (NFW) model in the inner regions, which gives a relation between the Hernquist scale length and the NFW scale radius, for a given concentration, as defined in Springel et al. (2005).  We consider models with a range of the circular velocity values at the virial radius, $v_{200}$, that correspond to a range of virial mass values, $M_{200}$, given in the literature.  We provide orbital distributions that span $v_{200}= 160 - 200$ km/s (which corresponds to $M_{200} = 1 - 1.8 \times 10^{12} M_{\odot}$), which spans the typical range of Milky Way masses found in the literature (Watkins et al. 2019; Deason et al. 2019; Posti \& Helmi 2019; Fritz et al. 2018; Piffl et al. 2014; Boylan-Kolchin et al. 2013). 

Given these initial conditions at $t = -1$ Gyr, we first use a parallelized implementation of the test particle code (Chang \& Chakrabarti 2011) to determine the range of disk response that corresponds to the orbital distribution determined from the Gaia proper motions and associated uncertainties.  We determine the initial conditions at $t = -1 $ Gyr because the current errors in the Gaia proper motions for Antlia 2 would not produce robust orbits for longer time integrations (Lipnicky \& Chakrabarti 2017).  The test particle calculations, which sample the errors in the proper motions, have been carried out for 3000 realizations for the $V_{200}=200$ km/s case. We then carry out a targeted set of SPH simulations with GADGET-2, as in earlier work (CB09).  The number of gas, stellar, and halo particles for our fiducial case are $8 \times 10^{5}, 8 \times 10^{5}, 1.2 \times 10^{6}$ respectively.  We have increased the number of particles in each component by a factor of two and find converged results for the metrics we use here.  The halo of the Milky Way is initialized with a Hernquist (1990) profile (matched to NFW in the inner regions) with an effective concentration of 9.39, a spin parameter $\lambda = 0.036$, and a range of circular velocities $V_{200}$ (see Table 1) that thereby correspond to a range of $M_{200}$ values. The simulated galaxies also include an exponential disc of stars and gas, with a flat extended H I disc, as found in surveys of spirals (e.g. Wong \& Blitz 2002).  The exponential disk size is set by requiring the disk mass fraction (taken to be 3.7 \% of the halo mass) is equal to the disk angular momentum, which results (for these parameters) in a disk scale length of 3.78 kpc.  The disk mass for the fiducial $v_{200} = 200$ km/s is $6.8 \times 10^{10} M_{\odot}$, and for the $v_{200} = 180$ km/s is $5 \times 10^{10} M_{\odot}$, which are comparable to observed values (Bovy \& Rix 2013).  For both cases, we assume 1:100 mass perturbers to represent Antlia 2's progenitor mass.

The simulated Antlia 2 dwarf galaxy is also similarly initialized, with stars and dark matter, but does not include gas.  Its concentration is set from relations derived from cosmological simulations, that show a correlation between the mass and concentration of dark matter halos (Maccio et al. 2008).  Antlia 2's progenitor mass is uncertain.  Its current stellar mass from its measured luminosity is $\sim 5\times 10^{5} M_{\odot}$ (Torrealba et al. 2019).  Given its mean [Fe/H] metallicity of -1.39, the Kirby et al. (2013) mass-metallicity relation would imply a stellar mass of $\sim 10^{7} M_{\odot}$.  The difference in the values of the current stellar mass and inferred stellar mass from the mass-metallicity relation may be due to tidal stripping of the satellite.  Using the $SFR-M_{200}$ relation of Erkal \& Read (2018), this would give a progenitor mass of $2 \times 10^{10} M_{\odot}$ for an age of 11.2 Gyr, where the age is as given in Torrealba et al. (2019).  Lower stellar masses of $\sim 10^{6} M_{\odot}$ would give $M_{200} \sim 3 \times 10^{9} M_{\odot}$.  Here, we consider $1:100$ mass-ratio progenitors for the Antlia 2 dwarf galaxy, which are roughly comparable to expectations from using the mass-metallicity relation, along with the $SFR-M_{200}$ relation.   

Comparisons with other dwarf galaxies also support a massive progenitor mass for Antlia 2.  Tucana is an isolated local group dSph. It has a stellar mass of $\sim 3 \times 10^{6}~M_{\odot}$, a lower metallicity than Antlia of [Fe/H] $\sim$ -1.95 and no evidence of significant stellar mass loss due to tides. Both abundance matching and its stellar kinematics favor a pre-infall halo mass of $M_{200} \sim 10^{10}~M_{\odot}$ (Gregory et al. 2019).  NGC6822 is an isolated dIrr with an estimated halo mass of $M200 \sim 2 \times 10^{10}~M_{\odot}$.    It has a present-day stellar mass of $7.6 \times 10^{7} M_{\odot}$, but has formed stars steadily for a Hubble time. If it stopped forming stars $\sim$ 11.2 Gyrs ago, its stellar mass (assuming a constant star formation rate) would have been $\sim 10^{7} M_{\odot}$ -- consistent with what we estimate for Antlia 2 given its high [Fe/H] (Read et al. 2016; Read et al. 2017).  (Note that [Fe/H] for NGC 6822 is [Fe/H] $\sim$ -1 which is higher than Antlia, as expected for its larger stellar mass.) Thus, Antlia 2 is roughly consistent with a "failed" NGC6822 that fell into the Milky Way $\sim$ 11 Gyrs ago.

Table 1 gives the parameters of the SPH simulations, including the simulation name, the $V_{200}$ and $M_{200}$ of the primary galaxy, the pericenter of the Antlia2 dwarf galaxy.   Here, we adopt an isothermal equation of state, which may be representative of the outskirts of galaxies where the energy injection from supernovae is low due to the low star formation rate in the outskirts (Bigiel et al. 2010).  

\section{Results}
\begin{figure}[ht]        
\begin{center}
\includegraphics[scale=0.48]{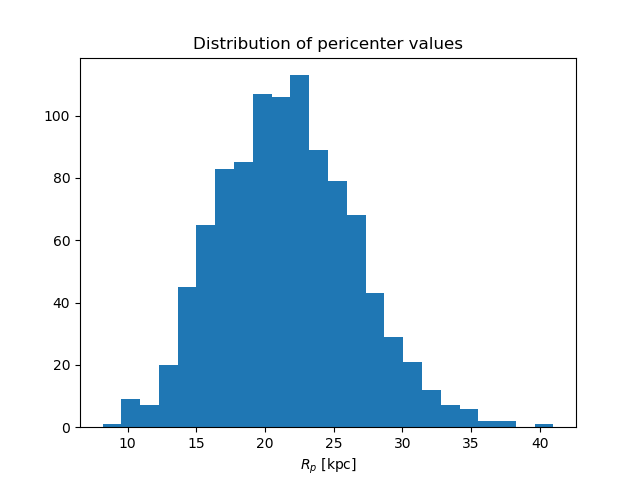}
\includegraphics[scale=0.48]{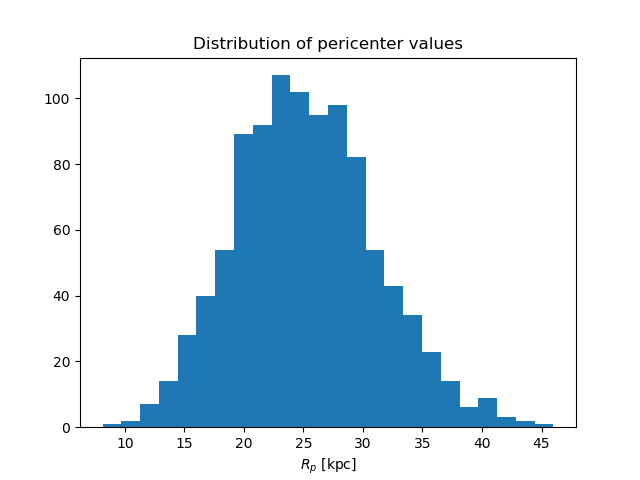}
\includegraphics[scale=0.48]{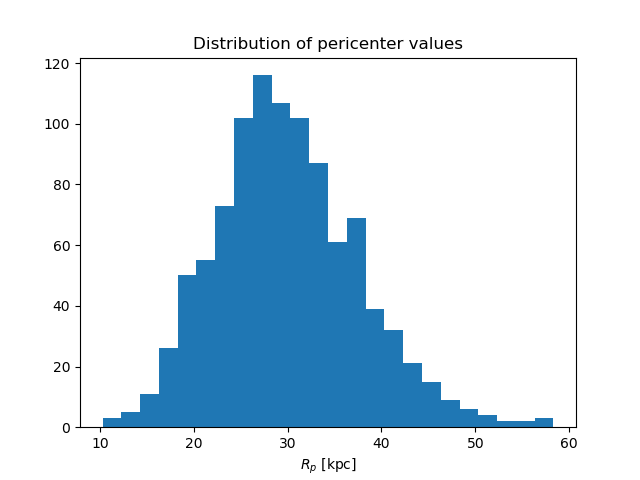}
\caption{(a) Pericenter distributions from Gaia PMs presented in Torrealba et al. (2019) for Antlia 2 for $v_{200} =$ 200 km/s (our fiducial model), (b) for $v_{200}$ =180 km/s,  (c) for $v_{200}$= 160 km/s \label{f:antorbits}}
\end{center}
\end{figure}

Figure \ref{f:antorbits} (a-c) shows Antlia 2's most recent pericenter distribution for $v_{200}$ = 200, 180 and 160 km/s, which vary in $M_{200}$ from $1.86 - 1 \times 10^{12} M_{\odot}$ from the backward time-integration of its orbits to $t=-1$ Gyr.  This range of MW masses is consistent with expectations from the literature, as noted in \S 2.  The mean of the pericenter distribution shrinks from 30 kpc for $M_{200} = 10^{12} M_{\odot}$ to 21 kpc for $M_{200} = 1.86\times 10^{12} M_{\odot}$ as the mass of the simulated MW increases, as expected.  However, these models all have a significant fraction of orbits with low pericenters given the 1 sigma errors in the proper motions reported by  Torrealba et al. (2019).  We have also carried out a similar exercise for the MW2014 potential that was employed by Torrealba et al. (2019), which is an adaptation from Bovy (2015) (but with a higher mass by a factor of two), and we find a mean pericenter of 38 kpc, with a tail of low pericenters extending to $\sim$ 10 kpc in that case also.  Our GADGET-2 and test particle calculations described below will employ the Hernquist-NFW potential. 


Given a projected surface density map, one can compute the individual $m-th$ Fourier amplitudes that describe the strength of the perturbing response as: 
\begin{equation}
a_{m}(r,t)= \frac{1}{2\pi}\int \Sigma(r,\phi) exp(-i m \phi) d\phi \\ ,
\end{equation}
where $\Sigma(r,\phi)$ is the projected gas surface density.  

The effective Fourier amplitude, $a_{m,eff}$ of the disk for an individual mode $m$ is then given by: 
\begin{equation}
a_{m,eff}(t) = \frac{1}{r_{\rm out} - r_{\rm in}}\int_{\rm r_{in}}^{\rm r_{out}} |a_{m}(r,t)| dr  \\ ,    
\end{equation}
where $r_{\rm in} = 10\,{\rm kpc}$ and $r_{\rm out} = 25\,{\rm kpc}$ are the inner and outer radii that we average over.
The quantity $a_{t,eff}(t)$ can be calculated by summing the effective response of the modes : 
\begin{equation}
a_{t,eff}(t) = \sqrt{\frac{1}{4} \sum_{m=1}^{m=4}|a_{m,eff}(t)|^{2}}     
\end{equation} 
In CC11 and in LCC18, we derived scaling relations for this quantity. Eqn 10 in LCC18 describes a fitting relation for $a_{t,eff}$ in terms of the ratio of the satellite mass ($m_{sat}$) to primary galaxy mass ($M_{host}$) and pericenter distance $R_{p}$ that we can use to roughly estimate the pericenter distance, given an assumed satellite mass to primary galaxy mass ratio, and an observed value for $a_{t,eff}$.  As discussed below, the observed HI data has wedges excised out of it, and so we have defined a new quantity, $a_{t,13}$, to mitigate its effects.  The observed HI data has a value of $a_{t,13} = 0.24$.  Using the relation defined in LCC18 as an estimate\footnote{We use this estimate (eqn.(10) of LCC18) under the assumption that the scaling for individual $a_{m,eff}$-modes scale similarly to $a_{t,eff}$.} for $a_{t,13}$ and using $m_{sat}/M_{host} = 1/100$, we obtain $R_{p} = 10~\rm kpc$. Thus, our rough expectation from scaling relations (LCC18) is that low pericenters would be ($\sim$ 10 kpc) needed to match the outer HI disk planar disturbances.  The scaling relations from LCC18 also indicate that the power in the Fourier modes scale as $(m_{sat}/M_{host})^{1/2}$.  Therefore, if the progenitor mass was $\sim 10^{9} M_{\odot}$, i.e., a 1:1000 mass ratio perturber, the power in the Fourier modes would be lower by a factor of 2.

The HI map constructed by LBH06 excludes regions that lie with $\pm$ 15 degrees of the Sun-Galactic center line because distances are difficult to determine in these regions as the velocity dispersion is larger than the line of sight velocity.  The wedges that are excised from the map will affect our calculations of the Fourier amplitudes.  Since the odd modes are less affected (Chakrabarti \& Blitz 2011) by the wedges, we focus here on the $m=1$ and $m=3$ modes, and our definition of $a_{t,13}$ will only include the sum of these modes, i.e.: 
\begin{equation}
  a_{\rm t, 13}(t) = \sqrt{\frac{1}{2} (|a_{m=1,eff}(t)|^{2} + |a_{m=3,eff}(t)|^{2})}     
\end{equation} 
where we sum (in quadrature) the $m=1$ and $m=3$ modes.  We first symmetrize the wedges, and these symmetrized wedges produce an artificial amount of power in even modes.  We have checked the effects of the angular cuts in our simulated data (test particle and SPH) and find that the power in the odd modes are not significantly affected by the (symmetrized) angular cuts.  


\begin{figure}[h]
\begin{center}
\includegraphics[scale=0.5]{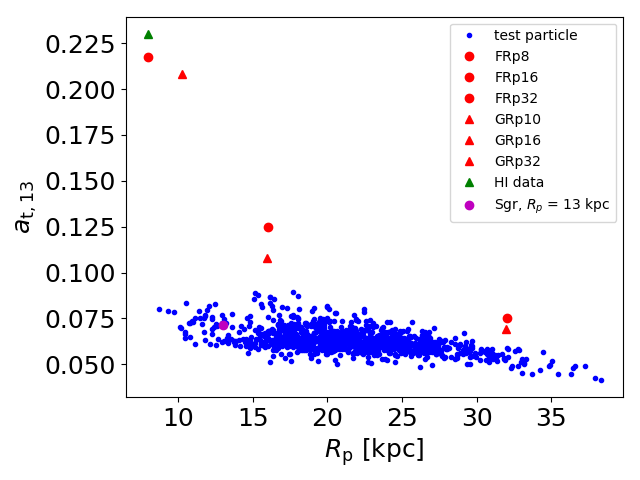}
\caption{Effective Fourier amplitudes vs pericenter, for test particle calculations (blue dots) that sample the uncertainty in the observed Gaia proper motions for $V_{200} = 200$ km/s, and SPH simulations of Antlia 2 (red) for specific realizations (see Table 1 for the description), along with the HI data (green) (shown at an arbitrary pericenter).  The Sgr dwarf case is also over-plotted in magenta, and its contribution is not sufficient to explain the disturbances in the outer HI disk of the Galaxy.  \label{f:ateff}}
\end{center}
\end{figure}

Figure \ref{f:ateff} depicts the effective Fourier amplitudes, $a_{t,13}$, from the test particle calculations (blue points) for our $V_{200} = 200$ km$\,{\rm s^{-1}}$ model, which samples the orbital distribution as defined above.  Red points are the results from our SPH calculations, and the HI data, in green, is shown at an arbitrary pericenter.  Here it is clear that only orbits with low pericenters ($R_{\rm p} \la 10$ kpc) are able to match the observed level of Fourier power in the outer HI disk of the Milky Way.  The red dots are our fiducial case ($v_{200} = 200$ km/s) and the red triangles are the $v_{200} = 180$ km/s case.  As expected (LCC18; CC11), $a_{t,13}$ primarily depends on $m_{sat}/M_{host}$ and the pericenter distance (the disk mass does not have a significant effect).  The test particle calculations underestimate the disk response relative to the SPH calculations, especially at low pericenters, due to the nature of the collisional gas in the SPH simulations. For larger pericenters, the results are quite similar, as would be expected.  Similar results are found for simulations done with only the stellar disk, which indicates that the self-gravity of the disk also influences the Fourier amplitudes.

We now investigate if the Sgr dwarf can excite the observed planar disturbances in the HI disk of the Galaxy.  We adopt a progenitor mass at $t = -1~\rm Gyr$ of $10^{10} M_{\odot}$, which is consistent with other models (Purcell et al. 2011; Laporte et al. 2017), accounting for the mass loss at $t=-1~\rm Gyr$ relative to the progenitor masses used at earlier times (several Gyr ago).  As with the Antlia2 dwarf galaxy, we derive the orbit distribution of the Sgr dwarf using the Gaia DR-2 proper motion (Helmi et al. 2018) combined with its radial velocity (McConnachie 2012) and an assumed heliocentric distance of $26~{\rm kpc}$ (Monaco et al. 2004). For the Sgr dwarf, the Gaia proper motions have very small errors, and therefore, for a given potential, its pericenter is tightly constrained. As shown in Figure \ref{f:ateff}, the Sgr dwarf (magenta point), does not drive sufficiently large planar disturbances to explain the observed HI data.  Sgr is on a polar orbit rather than Antlia 2's near-co-planar orbit, which we previously showed (CC11) is less effective in driving planar disturbances. Sgr's pericenter of $\sim$ 13 kpc is larger than the lowest pericenters of Antlia2's orbital distribution, which also leads to a reduced effect on the outer gas disk relative to what Antlia2 can produce, given the tail of low pericenters.  Finally, the time of pericenter also enters into this -- the minimum of $R_{p} = 13$ kpc occurs in Sgr's orbit at t=-0.05 Gyr, thus the effect we see now should be due to Sgr's previous disk crossing, which occurred at t = -0.4 Gyr, when the Sgr dwarf crossed the disk at 15 kpc.  This too is higher than the pericenters of $\sim$ 10 kpc for a 1:100 mass ratio perturber needed to match the power in the outer HI disk.  

\begin{figure}[h]
\begin{center}
\includegraphics[scale=0.5]{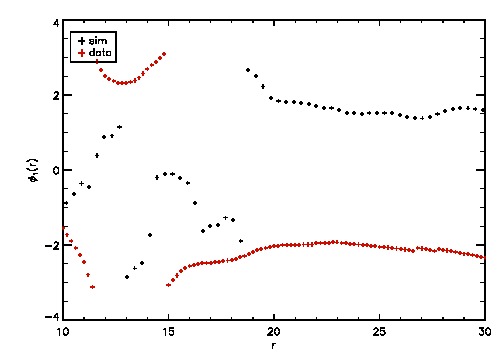}
\includegraphics[scale=0.38]{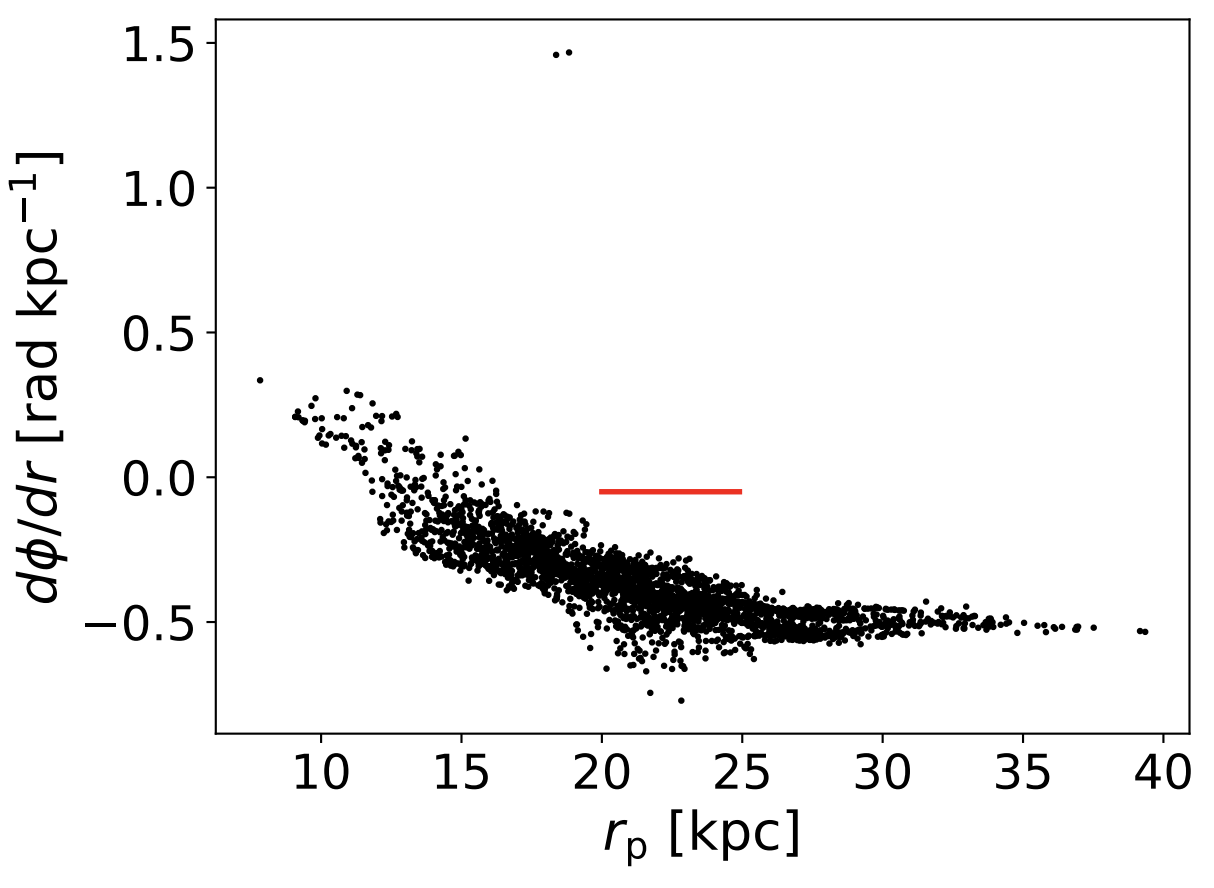}
\caption{(a) Phase of the m=1 mode vs radius from the FRp8 simulation (black), compared to the HI data (red), (b) $d\phi/dr$ of the m=1 mode vs r from test particle calculations, red line shows the gradient of the $m=1$ phase from the HI data. \label{f:phase}}
\end{center}
\end{figure}

The projected gas surface density can be decomposed into the Fourier amplitudes $a_{m}(r)$ and the phase of the modes $\phi_{m}(r)$.  The radial distribution of the phase of the modes expresses how tightly or loosely a spiral pattern is wrapped, and is given by: 
\begin{equation}
\phi(r,m) = arctan\frac{-Imag[FFT(\Sigma(r,\phi)]}{Re [FFT \Sigma(r,\phi)]}    
\end{equation}
Chakrabarti \& Blitz (2011) (henceforth CB11) used the phase of the modes to estimate the azimuthal location of the perturber.  We focus here on the $m=1$ phase, because the wedges in the raw HI map only minimally affect the odd modes as discussed above.  We ignored the $m=3$ mode as it has a three-fold degeneracy. Figure \ref{f:phase}(a) shows $\phi_{1}(r)$ from the FRp8 simulation that is able to match the observed disturbances in the outer HI disk of the MW (black), over-plotted with the phase from the raw HI data (red).  Interestingly, both the simulation and the data display a relatively flat phase variation in the outskirts.  A flat phase variation implies at at least within a certain radial range, the disturbances are nearly radial, which suggests that the origin of the observed HI disturbance cannot arise from a nonaxisymmetric perturbation at smaller radii, which would produce a tight spiral pattern, e.g. $d\phi_1/dr < 0$. Figure \ref{f:phase} (b) shows the average value of $d\phi_{1}/dr$ from 20 - 25 kpc from the test particle calculations (black) for a range of pericenters, over-plotted with the same metric from the raw HI data (red).  The low pericenter models display a relatively flat phase variation, while the larger pericenters have a negative slope.  Thus, the phase gives an independent constraint of the pericenter and more evidence favoring a subhalo excitation of the HI disk (CB09). 

\begin{figure}[h]
\begin{center}
\includegraphics[scale=0.55]{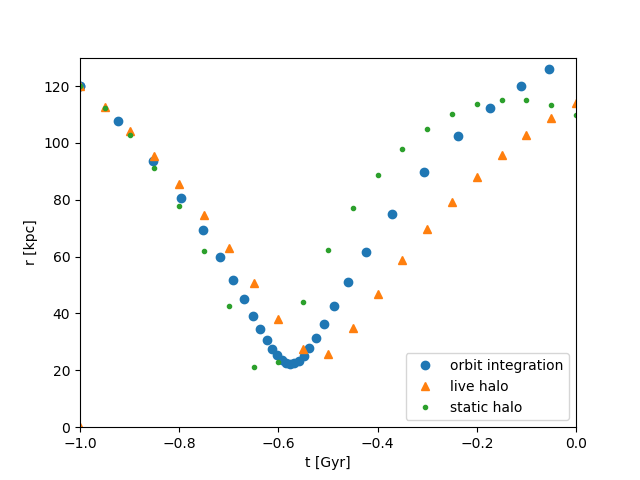}
\caption{Comparison of our orbit integration calculation over the past 1 Gyr (shown in blue dots), with our GADGET calculation with a live halo (orange triangles), and the corresponding calculation for a static halo (green dots).\label{f:orbitcomp}}
\end{center}
\end{figure}

We have restricted our calculations here to the past 1 Gyr due to the currently large error bars in the Gaia proper motion data; integrating back to even earlier times would result in significant uncertainties in the orbit, as shown in earlier work (Lipnicky \& Chakrabarti 2017; Lux et al. 2010).  Over this timescale, we have compared in Figure \ref{f:orbitcomp} the results of our orbit integration calculation (blue dots) to a GADGET calculation for Antlia2's orbit in a live halo (orange triangles), and to Antlia2's orbit in a static halo (green dots).  Although there are differences between these three calculations that arise from both tidal stripping of the Antlia 2 satellite over this timescale, the time evolution of the dark matter halo, and as well as our implementation of the Chandrasekhar dynamical friction formula (Chang \& Chakrabarti 2011), in all three cases, the difference is less than 20 \%, with the difference between the static halo case being generally less than 10 \%.   Similar results are found for the velocity comparison for these three difference kinds of calculations.  Thus, our analysis of Antlia 2's orbit here is approximate, and is not accurate to better than 20 \%.

\subsection{Proper motions from all of the Antlia 2 stars}

We have recalculated the proper motions of Antlia 2 using all of the stars identified as belonging to Antlia 2, using a more conservative background model.   Torrealba et al. (2019) determined the proper motion using only stars with radial velocities that they obtained follow-up observations of.  Here, we independently measure the mean proper motion of Antlia 2 by constructing a probabilistic model of the cluster and background stellar kinematics using astrometric and photometric data from Gaia. 
We start by selecting all Gaia sources within $2.5^\circ$ of the center of Antlia 2 with parallax $\varpi < 0.25~{\rm mas}$, then select candidate Antlia 2 member stars using photometric selections based on Figure 1 of Torrealba et al. (2019).
We then construct a probabilistic mixture model in proper motions for the foreground (i.e. Antlia 2) and background (i.e. Galactic field) using Gaussians in order to infer the mean proper motion of Antlia 2, following the approach described in Section 3.1 of Price-Whelan et al. (2018).
Briefly, we first mask the sky region immediately around Antlia 2 (removing stars within three times the effective radius defined in Torrealba et al. (2019) to define a ``background'' region, and construct a noise-deconvolved representation of the Galactic field proper motion distribution in this sky region using a five-component Gaussian mixture model.
We then represent Antlia 2 as an additional Gaussian component with unknown mean and variance.
We generate posterior samplings in the mean and internal proper motion dispersion of Antlia 2 and a mixture weight parameter ($f$ in Price-Whelan et al. (2018)) using an ensemble Markov Chain Monte Carlo (MCMC) sampler \texttt{emcee} (Foreman-Mackey et al. 2013). 
From this inference, which properly marginalizes over the unknown Galactic background properties and utilizes the full Gaia covariance matrix information for proper motions of stars in this field, we derive a proper motion for Antlia 2 of $(\mu_{\alpha} \cos\delta, \mu_{\delta}) = (-0.068, 0.032) \pm (0.023, -0.031)~\rm mas~\rm yr^{-1}$.

\begin{figure}[ht]        
\begin{center}
\includegraphics[scale=0.48]{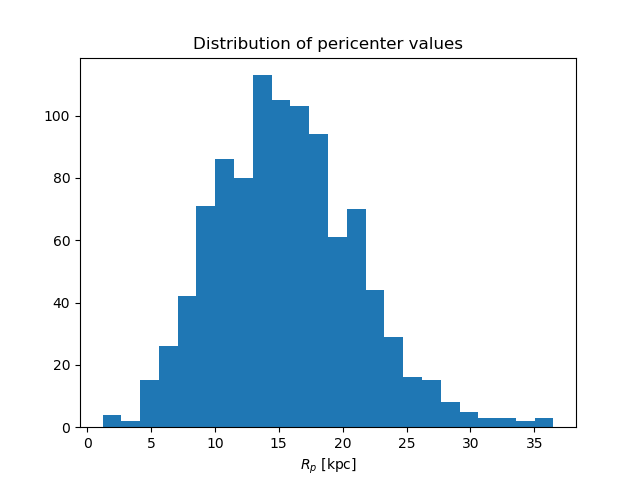}
\includegraphics[scale=0.48]{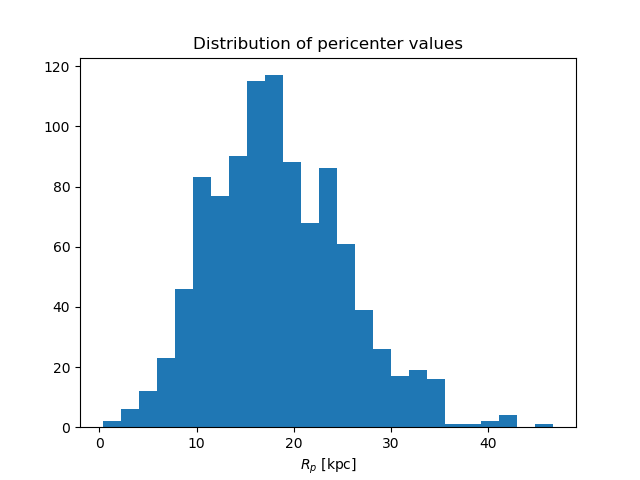}
\includegraphics[scale=0.48]{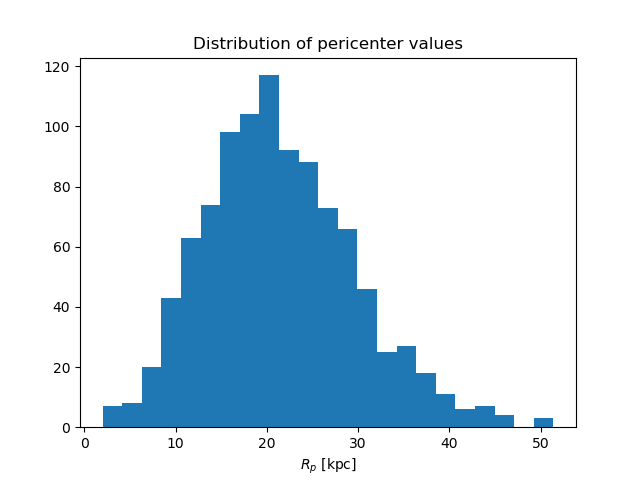}
\caption{(a) Pericenter distributions using all of the Antlia 2 stars to re-derive the proper motions (see text) for $v_{200} =$ 200 km/s (our fiducial model), (b) for $v_{200}$ =180 km/s,  (c) for $v_{200}$= 160 km/s \label{f:antorbitsAPW}}
\end{center}
\end{figure}

The revision in the mean value of the proper motions leads a revision in the pericenter distributions.  Figure \ref{f:antorbitsAPW} (a-c) shows the pericenter distributions for $v_{200} = 200$ km/s, 180 km/s and 160 km/s when we use all of the stars for Antlia 2.  These figures are analogous to Figure \ref{f:antorbits}, where we used the proper motions cited in Torrealba et al. (2019) to calculate the pericenter distributions.   One difference that is immediately clear is that upon using the proper motions derived from all of the Antlia 2 stars, the relative probability of the low pericenters increasing significantly for all the mass models, and is particularly significant for the $v_{200} = 200$ km/s case (where more than 70 \% of cases have low pericenters that are less than 10 kpc).  Thus this calculation shows that the low pericenters may very well be viable.  However, as with the proper motions cited in Torrealba et al. (2019), the errors in our proper motion inference are also large.  Therefore, this calculation should be redone upon the release of Gaia DR-3 data that will have considerably lower errors on the proper motions.

\section{Discussion \& Conclusion}

In summary, the orbital distributions for Antlia 2 have a tail of low pericenters of $\sim$ 10 kpc for a range of Milky Way masses commonly cited in the literature (from $\sim 10^{12} - 2 \times 10^{12} M_{\odot})$, when we employ the proper motions cited in Torrealba et al. (2019).  The probability of the low pericenters increases significantly when we employ our revised calculation of the proper motions using Gaia DR-2 data, where we have used all of the Antlia 2 stars with proper motions from Gaia DR-2.  

A close interaction of this kind with a 1:100 mass ratio perturber is sufficient to explain the planar disturbances observed in the outer HI disk of the Milky Way.  Moreover, the phase of the disturbances has a flat radial variation for the HI data, as do the Antlia 2 simulations with low pericenters, independently confirming that low pericenters are needed to match the disturbances manifest in the outer gas disk of the Galaxy.  We show that the tidal strength of the Sgr dwarf is insufficient to explain the disturbances in the outer gas of the Galaxy.  Of the other tidal players of the Milky Way, the Large and Small Magellanic Clouds are too distant and have not approached closer in the recent past (Besla et al. 2007; Besla et al. 2012) to account for this level of Fourier power in the outer HI disk.  Thus, Antlia 2 may be the driver of the observed large perturbations on the outskirts of our Galaxy. 

\begin{figure}[h]
\begin{center}
\includegraphics[scale=0.6]{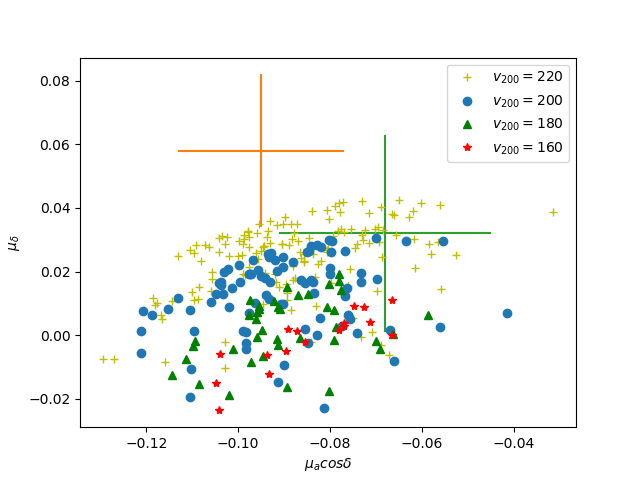}
\caption{The proper motions for Antlia2 (for $v_{200} =$ 160, 180, 200 and 220 km/s) that correspond to the low pericenters ($R_{p} \la 10$ kpc) of the orbital distributions. The Gaia proper motions and 1-sigma error given in Torrealba et al. (2019) are shown in orange, and our recalculation of the proper motions are shown in green.  Our independent calculation of the proper motion gives $(\mu_{\alpha} \cos\delta, \mu_{\delta}) = (-0.068, 0.032) \pm (0.023, -0.031)~\rm mas~\rm yr^{-1}$, which differs somewhat from that cited in Torrelaba et al. (2019).  If Antlia2 is indeed the dwarf galaxy that perturbed the outer HI disk, its $\mu_{\delta}$ values are constrained, while a large range of values for $\mu_{\alpha}cos \delta$ is possible (depending on the potential assumed).  \label{f:pmpredict}}
\end{center}
\end{figure}

If Antlia 2 is responsible for the outer HI disk planar disturbances, its proper motions are constrained to those that give orbits with low pericenters.  Figure \ref{f:pmpredict} shows the proper motions that correspond to the low pericenters of the orbital distribution $R_{p} \la$ 10 kpc for the $V_{200} = 220, 200, 180, 160$ km/s models, over-plotted with the mean and 1-sigma errors of the Gaia proper motions cited by Torrealba et al. (2019) (shown in orange), along with our calculation of the Gaia proper motions (shown in green).  The low pericenters correspond to proper motions with a wide range of $\mu_{\alpha} cos \delta$ values, but the $\mu_\delta$ values are constrained to be close to (and below) the Gaia proper motions cited by Torrealba et al. (2019). The $\mu_{\delta}$ values are nevertheless higher than kinematic proper motions cited by Torrealba et al. (2019).  Given that the probability of the low pericenters increases significantly when we use our proper motion calculations it is not surprising that our model predictions overlap with these revised proper motion values.  

Past HST measurements of proper motions for dwarf galaxies, such as the LMC, have changed significantly (and outside the scope of the 1-sigma errors) from the 2 epoch to 3 epoch values (Kallivayallil et al. 2013).  
The mean proper motion of the stars in the Antlia 2 dwarf galaxy are affected by correlated proper motion errors, and may well be revised upon future data releases.  Our prediction for Antlia 2's proper motions (for the potentials considered here), can soon be tested by upcoming improved data from Gaia DR-3 and Gaia DR-4. Possibly machine learning techniques may be able to improve on the proper motion measurement errors even before the release of Gaia DR-3 and DR-4 data. 

Antlia 2 presents an unique laboratory for the study of a dark-matter dominated dwarf galaxy, if it is indeed the perturber that drove the ripples in the outer gas disk of our Galaxy.  Since its mass was \emph{predicted} from a dynamical analysis, its effect on the Galaxy sets bounds on its dark matter content more strictly than forward-modeling approaches.  With its half-light radius of 3 kpc, one may also be able to obtain more stringent constraints on its dark matter density profile than for other dwarf galaxies, and thereby effectively discriminate between self-interacting dark matter and CDM models (Fry et al. 2015).  One mechanism that may explain the diversity of dark matter density profiles in Milky Way dwarf galaxies are "dark matter heating" models (Read et al. 2019).  In a forthcoming companion paper, we investigate the structure of Antlia 2, contrasting the effects of self-interacting dark matter models and CDM models on the evolution of its density profile.  Kahlhoefer et al. (2019) have recently noted that self-interacting dark matter models with large cross sections may help to explain the diversity of density profiles in Milky Way dwarf galaxies, from very compact systems like Draco to very diffuse systems like CraterII, especially when their orbital evolution is considered, as the time evolution of the density profile depends sensitively on the orbit of the dwarf galaxy.  

\bigskip
\bigskip

\acknowledgments

Much of this work was conducted at the KITP Santa Barbara long-term program "Dynamical Models for Stars and Gas in Galaxies in the Gaia Era".  SC gratefully acknowledges the hospitality of the KITP during her visit.  The simulations have been performed on Xsede, NERSC, and on Google Cloud.  SC gratefully acknowledges support from NASA ATP NNX17AK90G, NSF AAG grant 1517488, and from Research Corporation for Scientific Advancement's Time Domain Astrophysics Scialog. PC is supported by NASA grant NNH17ZDA001N-ATP, NSF CAREER grant AST-1255469, and the Simons Foundation. This research was supported in part by Grant No. NSF PHY-1748958. We also acknowledge the Flatiron Institute for providing HPC resources that have contributed to the results reported
here.

\end{document}